\documentclass[aps,prc,reprint,amsmath,nofootinbib]{revtex4-1}

\usepackage[utf8]{inputenc}
\usepackage{amsmath}
\usepackage{amssymb}
\usepackage{multirow}
\usepackage[inline]{enumitem}

\usepackage{color}
\definecolor{theblue}{RGB}{0,50,230}

\usepackage{hyperref}
\hypersetup{
  colorlinks=true,
  linkcolor=theblue,
  citecolor=theblue,
  urlcolor=theblue
}

\usepackage{graphicx}
\graphicspath{{fig/}}

\newcommand{\trento}{T\raisebox{-0.5ex}{R}ENTo}
\newcommand{\avg}[1]{\langle #1 \rangle}
\newcommand{\nch}{N_\text{ch}}
\newcommand{\npart}{N_\text{part}}
\newcommand{\sqrts}{\sqrt{s_{NN}}}
\newcommand{\T}{\tilde{T}}
\newcommand{\Qs}[1]{Q_{s,\text{#1}}}
\newcommand{\vnk}[2]{v_#1\{#2\}}
\newcommand{\paddedhline}{\noalign{\smallskip}\hline\noalign{\smallskip}}
\newcommand{\order}[1]{$\mathcal O(10^{#1})$}

\usepackage{xparse}
\NewDocumentCommand\fig{sO{t}mm}{
  \begin{figure\IfBooleanT{#1}{*}}[#2]
    \includegraphics{#3}
    \caption{\label{fig:#3}#4}
  \end{figure\IfBooleanT{#1}{*}}
}

\begin{document}

\title{
  Applying Bayesian parameter estimation to relativistic heavy-ion collisions: \\
  simultaneous characterization of the initial state and quark-gluon plasma medium
}

\author{Jonah E.\ Bernhard}
\author{J.\ Scott Moreland}
\author{Steffen A.\ Bass}
\affiliation{Department of Physics, Duke University, Durham, NC 27708-0305}

\author{Jia Liu}
\author{Ulrich Heinz}
\affiliation{Department of Physics, The Ohio State University, Columbus, OH 43210-1117}

\date{\today}

\begin{abstract}
  We quantitatively estimate properties of the quark-gluon plasma created in ultra-relativistic heavy-ion collisions utilizing Bayesian statistics and a multi-parameter model-to-data comparison.
  The study is performed using a recently developed parametric initial condition model, \trento, which interpolates among a general class of particle production schemes, and a modern hybrid model which couples viscous hydrodynamics to a hadronic cascade.
  We calibrate the model to multiplicity, transverse momentum, and flow data and report constraints on the parametrized initial conditions and the temperature-dependent transport coefficients of the quark-gluon plasma.
  We show that initial entropy deposition is consistent with a saturation-based picture, extract a relation between the minimum value and slope of the temperature-dependent specific shear viscosity, and find a clear signal for a nonzero bulk viscosity.
\end{abstract}

\maketitle

\section{Introduction}

Simulations based on relativistic viscous hydrodynamics have been highly successful describing a wealth of bulk observables in heavy-ion collisions at the Relativistic Heavy-Ion Collider (RHIC) in Brookhaven, NY and the Large Hadron Collider (LHC) in Geneva, Switzerland.
Initially, the success of hydrodynamic simulations was primarily qualitative.
The framework elegantly described a number experimental phenomena, for example the existence of large azimuthal particle correlations, the mass ordering of these correlations, and their characteristic momentum dependence.

Modern hydrodynamic simulations have greatly expanded upon the successes of first-generation models.
The addition of dissipative corrections to ideal hydrodynamics \cite{Muronga:2004sf, Chaudhuri:2006jd, Romatschke:2007mq, Dusling:2007gi, Song:2007ux, Luzum:2008cw}, event-by-event fluctuations in the colliding nuclei \cite{Alver:2008zza, Alver:2010gr}, and modern lattice quantum chromodynamics (QCD) calculations for the quark-gluon plasma (QGP) equation of state \cite{Bazavov:2009zn, Borsanyi:2013bia, Bazavov:2014pvz} are just a few examples of developments which have dramatically improved the agreement of hydrodynamic models with experiment.

These developments have positioned hydrodynamic modeling to evolve beyond a qualitative science and quantitatively extract intrinsic properties of hot and dense QCD matter.
A primary goal of the ongoing effort is to determine the temperature dependence of QGP transport coefficients such as the specific shear viscosity $\eta/s$, theorized to reach a lower bound ${\eta/s \ge 1/4\pi}$ near the QGP phase transition temperature \cite{Danielewicz:1984ww, Policastro:2001yc, Kovtun:2004de}.
An estimate of the effective (constant) QGP shear viscosity needed to fit spectra and flows at RHIC found ${1 \le 4\pi\eta/s \le 2.5}$ \cite{Song:2010mg}, while independent studies have reported estimates consistent with this range \cite{Luzum:2008cw,Schenke:2010rr,Luzum:2012wu}.

The remaining uncertainty in $\eta/s$ arises largely from the hydrodynamic initial conditions:
different initial condition models lead to different hydrodynamic flow and hence prefer different values of $\eta/s$.
Current efforts to reduce uncertainties include improving theoretical descriptions of the initial conditions \cite{Schenke:2012wb, Niemi:2015qia} and testing respective model predictions against sensitive new observables \cite{Aad:2013xma, Aad:2014fla, ALICE:2016kpq}.
The process thus defines an iterative cycle in which theory calculations are embedded in hydrodynamic transport simulations, analyzed against a comprehensive list of bulk observables, and used to generate testable predictions which inform subsequent refinements to the theory.

Model optimization and comparison is often complicated by multiple undetermined and highly correlated input parameters.
In addition to QGP transport coefficients, simulations depend on auxiliary inputs such as an effective nucleon width and QGP thermalization time, all of which must be simultaneously optimized.
Evaluating a model for a single set of parameters requires thousands of individual event simulations, so direct optimization techniques quickly become intractable.

One solution to the model optimization problem is the use of modern Bayesian methods to estimate the parameters of computationally intensive models \cite{OHagan:2006ba, Higdon:2008cmc, Higdon:2014tva, Wesolowski:2015fqa}.
A given model is first evaluated at a relatively small number of parameter configurations and the results are interpolated by a Gaussian process emulator \cite{Rasmussen:2006gp}.
Then, using the emulator as a stand-in for the full model, a standard Markov chain Monte Carlo (MCMC) algorithm exhaustively explores the parameter space and extracts probability distributions for the optimal values of each parameter.

Bayesian methods have been applied to heavy-ion collisions in several previous studies \cite{Petersen:2010zt, Novak:2013bqa, Pratt:2015zsa, Sangaline:2015isa, Bernhard:2015hxa, Heinz:2015arc}, including simulations initialized with a two-component Monte Carlo Glauber (MC-Glb.) model \cite{Miller:2007ri} and the Kharzeev-Levin-Nardi (MC-KLN) model \cite{Drescher:2006pi}, an implementation of color glass condensate (CGC) effective field theory \cite{Gelis:2010nm, Gelis:2015gza}.
Future work could expand this coverage to additional calculations of QGP initial conditions in order to systematically constrain each model's parameters along with hydrodynamic transport coefficients.
Once the models are appropriately optimized, the relative accuracy of the various theory calculations may be quantified using a model selection criterion such as Bayes factors.

An alternative approach to model-by-model validation is to optimize parametric initial conditions that are sufficiently flexible to mimic the behavior of various theory calculations.
This allows the parameter optimization process to determine the nature of the initial conditions concurrently with QGP medium properties while propagating any relevant uncertainties---without imposing the assumptions of a specific model.
It also accelerates the model evaluation cycle by establishing which theory calculations are most compatible with the data and informing further refinements.
To this end, several recent studies have successfully used event-averaged parametric initial conditions to constrain QGP properties including the equation of state \cite{Novak:2013bqa, Pratt:2015zsa, Sangaline:2015isa}.

In this work, we extend previous efforts to parametrize and constrain QGP initial conditions using a recently developed event-by-event model, \trento\ \cite{Moreland:2014oya}, which is constructed to interpolate a subspace of all initialization models including (but not limited to) specific calculations in CGC effective field theory.
We couple the parametric model to viscous hydrodynamics and a hadronic afterburner and apply Bayesian methods to simultaneously estimate QGP initial condition and medium properties.

\section{Evolution model}

Heavy-ion collision dynamics are modeled in a multi-stage approach using relativistic viscous hydrodynamics for the time evolution of the QGP medium and microscopic Boltzmann transport to simulate the dynamics of the system after hadronization.

\subsection{Hydrodynamics and Boltzmann transport}

Relativistic hydrodynamics models calculate the time evolution of the QGP medium via the conservation equations
\begin{equation}
  \partial_\mu T^{\mu\nu} = 0
  \label{eq:conservation}
\end{equation}
for the energy-momentum tensor
\begin{equation}
  T^{\mu\nu} = e \, u^\mu u^\nu  - \Delta^{\mu\nu} (P + \Pi) + \pi^{\mu\nu},
\end{equation}
provided a set of initial conditions for the fluid flow velocity $u^\mu$, energy density $e$, pressure $P$, shear stress tensor $\pi^{\mu\nu}$, and bulk viscous pressure $\Pi$.
We use VISH2+1 \cite{Song:2007ux}, a stable, extensively tested implementation of boost-invariant viscous hydrodynamics which has been updated to handle fluctuating event-by-event initial conditions \cite{Shen:2014vra} and incorporate bulk viscous corrections with shear-bulk coupling \cite{Liu:2015bik}.
This implementation calculates the time evolution of the viscous corrections through the second-order Israel-Stewart equations \cite{Israel:1979wp, Israel:1976aa} in the 14-momentum approximation, which yields a set of relaxation-type equations \cite{Denicol:2014vaa, Ryu:2015vwa}
\begin{subequations}
  \label{eq:relaxation}
  \begin{align}
    \tau_\Pi \Pi + \dot{\Pi} &=
      - \zeta \theta - \delta_{\Pi\Pi} \Pi\theta
      + \lambda_{\Pi\pi} \pi^{\mu\nu} \sigma_{\mu\nu}, \\[1ex]
    \tau_\pi \dot{\pi}^{\langle \mu\nu \rangle} + \pi^{\mu\nu} &=
      2\eta\sigma^{\mu\nu} - \delta_{\pi\pi} \pi^{\mu\nu} \theta
      + \phi_7 \pi_\alpha^{\langle \mu} \pi^{\nu \rangle \alpha} \nonumber \\
      &\qquad {} - \tau_{\pi\pi} \pi_\alpha^{\langle \mu}\sigma^{\nu \rangle \alpha}
      + \lambda_{\pi\Pi} \Pi \sigma^{\mu\nu}.
  \end{align}
\end{subequations}
Here, $\eta$ and $\zeta$ are the shear and bulk viscosities, parametrized below.
For the remaining transport coefficients, we use analytic results derived in the limit of small but finite masses \cite{Denicol:2014vaa}.

The hydrodynamic equations of motion must be closed by an equation of state (EoS), $P = P(e)$.
We use a modern QCD EoS based on continuum extrapolated lattice calculations at zero baryon density published by the HotQCD collaboration \cite{Bazavov:2014pvz} and blended into a hadron resonance gas EoS in the interval {$110 \le T \le 130$~MeV} using a smoothstep interpolation function \cite{Moreland:2015dvc}.
The HotQCD EoS, characterized by the parametrized interaction measure $(e - 3P)/T^4$, has been compared to additional state-of-the-art calculations by the Wuppertal-Budapest collaboration and shown to agree within published errors \cite{Bazavov:2014pvz}.
The two parametrizations were also studied in a recent error analysis at RHIC energies which quantified the effect of systematic lattice EoS discrepancies and statistical continuum extrapolation errors on hydrodynamic observables \cite{Moreland:2015dvc}.
The effect of these errors on mean $p_T$, elliptic flow $v_2$, and triangular flow $v_3$ was found to be $\mathcal O(1\%)$ and hence is expected to be negligible in the present analysis.

In order to estimate the shear and bulk viscosities, we parametrize their temperature dependence and define several variable model inputs.
The viscosities are typically expressed as dimensionless ratios $\eta/s$ and $\zeta/s$, where $s$ is the entropy density;
for the specific shear viscosity $\eta/s$, we use a piecewise linear parametrization
\begin{equation}
  (\eta/s)(T) =
  \begin{cases}
    (\eta/s)_\text{min} + (\eta/s)_\text{slope} (T - T_c) & T > T_c \\
    (\eta/s)_\text{hrg}                                   & T \le T_c
  \end{cases},
  \label{eq:etas}
\end{equation}
motivated by calculations in low- and high-temperature limits which demonstrate that $\eta/s$ has a minimum near the QCD transition temperature \cite{Prakash:1993bt, Arnold:2003zc, Csernai:2006zz}.
We fix the transition temperature $T_c = 0.154$~GeV to match the HotQCD EoS \cite{Bazavov:2014pvz} and leave $(\eta/s)$ hrg, min, and slope as tunable parameters, with the slope restricted to non-negative values.
For the specific bulk viscosity $\zeta/s$, we use the parametrization \cite{Denicol:2009am, Ryu:2015vwa}
\begin{equation}
  (\zeta/s)(T) =
  \begin{cases}
    \begin{aligned}
      C_1 &+ \lambda_1 \exp [(x-1)/\sigma_1]  \\ &+ \lambda_2 \exp [ (x-1)/\sigma_2]
    \end{aligned}
    &T < T_a \\[3ex]
    A_0 + A_1 x + A_2 x^2 &T_a \le T \le T_b \\[2ex]
    \begin{aligned}
      C_2 &+ \lambda_3 \exp [-(x-1)/\sigma_3]  \\ &+ \lambda_4 \exp [-(x-1)/\sigma_4]
    \end{aligned}
    &T > T_b
  \end{cases},
  \label{eq:zetas}
\end{equation}
with $x = T/T_0$ and coefficients
\begin{align*}
  &C_1=0.03,\quad C_2=0.001, \\
  &A_0=-13.45,\quad A_1=27.55,\quad A_2=-13.77, \\
  &\sigma_1=0.0025,\quad \sigma_2=0.022,\quad \sigma_3=0.025,\quad \sigma_4=0.13, \\
  &\lambda_1=0.9,\quad \lambda_2=0.22,\quad \lambda_3=0.9,\quad \lambda_4=0.25, \\
  &T_0 = 0.18 \text{ GeV},\quad T_a = 0.995\, T_0,\quad T_b = 1.05\, T_0.
\end{align*}
Qualitatively, this form peaks near $T_0 = 180$~MeV and falls off exponentially on either side.
To estimate the magnitude of bulk viscosity, we scale $(\zeta/s)(T)$ by a tunable overall normalization factor $(\zeta/s)_\text{norm}$.

As the hydrodynamic medium expands and cools below the QCD transition temperature $T_c$, it undergoes a transition from a deconfined QGP to a hadron resonance gas (HRG).
We therefore convert the medium to an ensemble of particles and switch from hydrodynamics to a microscopic kinetic model, which can better handle the late stages of the collision including species-dependent kinetic freeze-out, hadronic feed-down dynamics, and non-equilibrium breakup.
Kinetic models also naturally account for hadronic viscosity, obviating the need to manually specify transport coefficients.
Thus, although the parametrizations for $\eta/s$ and $\zeta/s$, Eq.~\eqref{eq:etas} and \eqref{eq:zetas}, extend below $T_c$, they do not affect the kinetic model.
In particular, the parameter $(\eta/s)_\text{hrg}$ only controls the small fraction of hydrodynamic evolution below $T_c$ and before switching to the kinetic model, and hence is not expected to strongly affect the overall model.
Such multi-stage approaches are known as hybrid models \cite{Bass:2000ib, Nonaka:2006yn, Petersen:2008dd}.

\newcommand{\Tsw}{$T_\text{switch}$}

The conversion to particles, or ``particlization'', is performed on an isothermal spacetime hypersurface defined by a pre-specified switching temperature \Tsw.
Particlization denotes the conversion of the hadronic medium from macroscopic to microscopic degrees of freedom---distinct from the physical hadronization process---and in principle, may occur at any temperature within a small window near the QCD transition temperature, within which both the hydrodynamic and microscopic models predict the same medium evolution.
To test this postulate, we leave \Tsw\ as a variable parameter.
As the hydrodynamic medium cools past the switching temperature, particles are sampled from the Cooper-Frye formula \cite{Cooper:1974mv}
\begin{equation}
  E \frac{dN_i}{d^3p} =
    \frac{g_i}{(2\pi)^3} \int_\Sigma f_i(x,p) \, p^\mu \, d^3\sigma_\mu,
  \label{eq:cooper_frye}
\end{equation}
where $i$ is an index over species, $f_i$ the particle species' distribution function, and $d^3\sigma_\mu$ a volume element (located at spacetime position $x$) of the isothermal hypersurface $\Sigma$ defined by \Tsw.
We use the iSS sampler \cite{Shen:2014vra, Qiu:2013wca} for particlization.

\newcommand{\df}{\delta f}

The distribution function $f$ includes any non-equilibrium contributions from shear and bulk viscosities, typically expanded into an ideal part and a viscous correction, $f = f_0 + \df$, where the ideal part $f_0$ is a Bose or Fermi distribution and the viscous correction $\df = \df_\text{shear} + \df_\text{bulk}$.
We use a common form for the shear correction \cite{Teaney:2003kp}
\begin{equation}
  \delta f_\text{shear} = f_0(1 \pm f_0) \frac{1}{2T^2(e+P)} p^\mu p^\nu \pi_{\mu\nu}.
\end{equation}
The bulk viscous correction has a variety of proposed forms, each of which predicts significantly different behavior when either the bulk pressure $\Pi$ or momentum $p$ are large \cite{Dusling:2011fd, Noronha-Hostler:2013gga}.
Given this uncertainty and the small $\zeta/s$ at particlization [see Eq.~\eqref{eq:zetas}], we assume that bulk corrections will be small and neglect them for the present study, i.e.\ $\df_\text{bulk} = 0$.
This precludes any quantitative conclusions on bulk viscosity, since we are only allowing bulk viscosity to affect the hydrodynamic evolution, not particlization.
We will, however, be able to determine whether $\zeta/s$ is nonzero.
We plan to remedy this shortcoming in future work, enabling a quantitative estimate of the temperature dependence of bulk viscosity.

Once the fluid is converted into hadrons, the subsequent microscopic dynamics are simulated using the Ultra-relativistic Quantum Molecular Dynamics (UrQMD) model as a hadronic afterburner \cite{Bass:1998ca, Bleicher:1999xi}.
UrQMD uses Monte Carlo techniques to solve the Boltzmann equation
\begin{equation}
  \frac{df_i(x,p)}{dt} = \mathcal{C}_i(x, p),
\end{equation}
where $f_i$ is the distribution function and $\mathcal{C}_i$ the collision kernel for particle species $i$.
The model propagates all produced hadrons along classical trajectories, and accounts for their scattering, resonance formation, and decay processes until all hadrons in the system have ceased interacting.
The final particle data are then postprocessed into observables for comparison with experiment.

\subsection{Parametric initial conditions}

The hydrodynamic equations of motion necessitate initial conditions for the energy density $e$, fluid flow velocity $u^\mu$, shear stress tensor $\pi^{\mu\nu}$, and bulk pressure $\Pi$ at time $\tau = \tau_0$, when the system is assumed to have thermalized.
These initial conditions emerge from dynamical processes of the collision, and are commonly modeled in two stages: initial state models describe the system immediately after impact at time $\tau=0^+$, then pre-equilibrium transport models evolve the system until the thermalization time $\tau_0$.
Efforts to realistically model the pre-equilibrium stage include transport dynamics \cite{Schenke:2012wb, Schenke:2012fw, vanderSchee:2013pia, vanderSchee:2015rta, Chesler:2015fpa} motivated by thermalization studies in strong and weakly coupled field theories \cite{Romatschke:2006nk, Krasnitz:2002ng, Berges:2013eia, Arnold:2004ti, Kurkela:2011ti, Heller:2012km, vanderSchee:2013pia, Rebhan:2004ur, Heller:2011ju, Janik:2006gp}.

The importance of pre-equilibrium dynamics was recently studied by initializing hydrodynamic simulations with a free streaming phase (zero coupling) and switching to hydrodynamics (strong coupling) after different periods of time \cite{Liu:2015nwa, Heinz:2015arc}.
The authors showed that although free streaming never leads to thermalization, it can be used to bracket the influence of pre-equilibrium dynamics on the medium evolution as the pre-equilibrium coupling strength is expected to fall between the free streaming and hydrodynamic limits.
When bulk viscous effects were neglected, the analysis found a preference for a brief free streaming phase $\tau_\text{fs} \lesssim 1$~fm/$c$, but the effect on hydrodynamic bulk observables was small and modifications to the preferred value of the QGP specific shear viscosity $\eta/s$ were less than 10\%.
Including nonzero bulk viscosity opened a window for a longer free-streaming stage with $\tau_\mathrm{fs} \approx 2$~fm/$c$ and reduced the best-fit value for the specific shear viscosity by 20\%.
In real situations where the pre-equilibrium coupling strength is necessarily nonzero, dynamical effects on the extracted transport coefficients are expected to be even smaller.

In the present study we neglect pre-equilibrium dynamics, instead initializing the flow velocity to zero as well as the viscous terms, which quickly relax to their Navier-Stokes values \cite{Song:2009rh}.
This reduces the initial conditions to a thermal energy density, which may be provided as an entropy density and converted via the QCD EoS.
We generate event-by-event initial conditions using the recently developed parametric model \trento\ \cite{Moreland:2014oya}.
The model begins with a standard Monte Carlo Glauber formalism, summarized below, and parametrizes entropy deposition as a function of local participant nuclear density.

First, nucleon positions for nuclei $A$ and $B$ are sampled from a standard uncorrelated Woods-Saxon distribution \cite{Loizides:2014vua} and shifted by $\pm b/2$, where $b$ is a minimum-bias impact parameter.
Participants are then determined randomly from the inelastic collision probability \cite{d'Enterria:2010hd}
\begin{equation}
  \begin{aligned}
    P_\text{coll}(b) &= 1 - \exp\bigl[ -\sigma_{gg} T_{pp}(b) \bigr], \\
    T_{pp}(b) &= \int dx \, dy \, T_p(x, y) T_p(x - b, y),
  \end{aligned}
\end{equation}
where $b$ is now the impact parameter between two nucleons, $T_p$ is the nucleon thickness function, and the effective partonic cross section $\sigma_{gg}$ is fixed to reproduce the inelastic nucleon-nucleon cross section
\begin{equation}
  \sigma_\text{NN}^\text{inel} = \int 2 \pi b\, db\, P_\text{coll}(b).
\end{equation}
The energy-dependent cross section $\sigma^\text{inel}_\text{NN} = 4.0$, 4.2, 6.4, 7.0~fm$^2$ at $\sqrts=130$, 200, 2760, 5020~GeV, respectively \cite{Adare:2015bua, ATLAS:2011ag, ALICE:2012xs}.
For the nucleon thickness function we use a Gaussian
\begin{equation}
  T_p(x, y) = \frac{1}{2\pi w^2} \exp\bigg(\!-\frac{x^2 + y^2}{2 w^2}\bigg),
  \label{eq:Tp}
\end{equation}
where $w$ is a tunable effective nucleon width.

We now define the \emph{participant} thickness function
\begin{equation}
  \T(x, y) = \sum\limits_{i=1}^{N_\text{part}} \gamma_i\, T_p(x - x_i, y - y_i),
  \label{eq:participant}
\end{equation}
which differs from the conventional thickness function $T$ by including only participant nucleons and weighting each participant by a random factor $\gamma_i$, sampled from a gamma distribution with unit mean and variance $1/k$, where $k$ is a tunable shape parameter \cite{Bozek:2013uha}.
These weights are inserted to account for minimum-bias proton-proton multiplicity fluctuations.

\fig{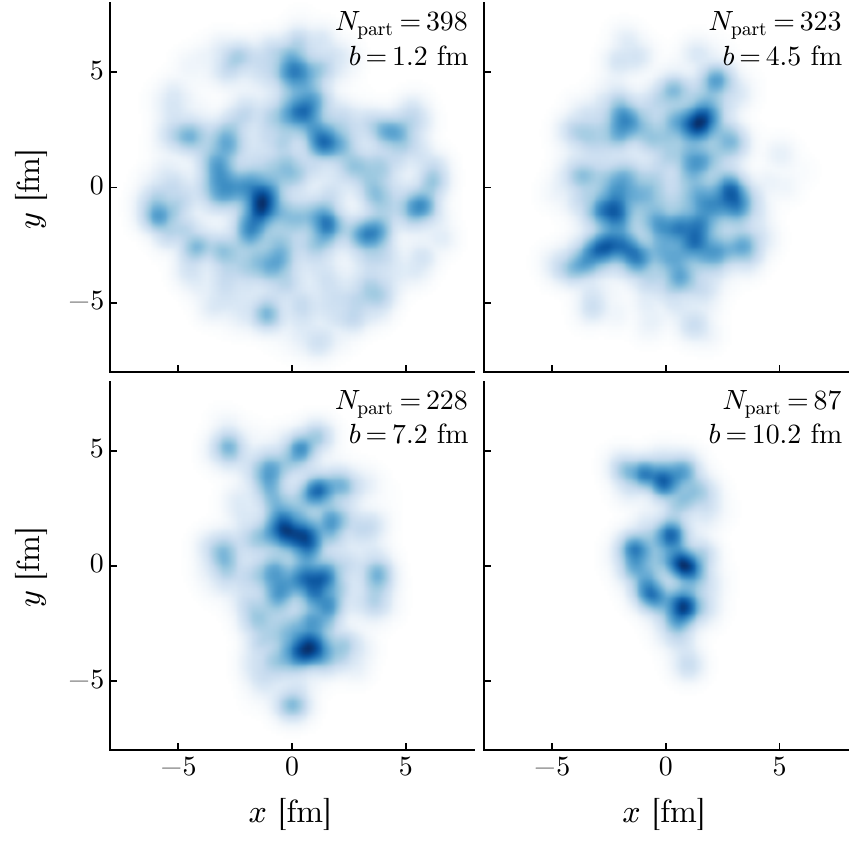}{
  Several randomly generated \protect\trento\ Pb+Pb initial condition events using generalized mean parameter $p=0$, nucleon width $w=0.5$~fm, and gamma fluctuation factor $k=1.4$.
}

The \trento\ model calculates local entropy density at midrapidity by applying a function $f$ to the participant thickness functions:
\begin{equation}
  s(\tau_0, x, y)\vert_{\eta_s=0} = f(\T_A, \T_B).
  \label{eq:mapping}
\end{equation}
We use a functional form motivated by basic physical constraints and phenomenological observations \cite{Moreland:2014oya} known as the generalized mean:
\begin{equation}
  s \propto \left( \frac{\T_A^p + \T_B^p}{2} \right)^{1/p}.
  \label{eq:genmean}
\end{equation}
This parametrization introduces a continuous entropy deposition parameter $p$ which effectively interpolates among different entropy deposition schemes.
For ${p=(1, 0, -1)}$, the generalized mean reduces to arithmetic, geometric, and harmonic means, while for ${p \rightarrow \pm\infty}$ it asymptotes to minimum and maximum functions:
\begin{equation}
  \newlength{\extraspace}
  \setlength{\extraspace}{0.5ex}
  s \propto
  \begin{cases}
    \max(\T_A, \T_B) & p \rightarrow +\infty, \\[\extraspace]
    (\T_A + \T_B)/2 & p = +1, \hfill \text{ (arithmetic)} \\[\extraspace]
    \sqrt{\T_A \T_B} & p = 0, \hfill \text{ (geometric)} \\[\extraspace]
    2\, \T_A \T_B/(\T_A + \T_B) & p = -1, \hfill \text{ (harmonic)} \\[\extraspace]
    \min(\T_A, \T_B) & p \rightarrow -\infty.
  \end{cases}
  \label{eq:means}
\end{equation}
Perhaps the simplest explanation of this ansatz is to examine the effect of the mapping on realistic events:
Fig.~\ref{fig:trento_events} shows examples of entropy density in the transverse plane for several typical Pb+Pb events at $\sqrts=2.76$~TeV,
while Fig.~\ref{fig:thickness} shows a cross section of a single event along the direction of the impact parameter.
At each point in the transverse plane there are two relevant scales of interest:
the smaller of the two participant densities, $\T_\text{min} = \min(\T_A, \T_B)$, and the larger, $\T_\text{max}$.
In Fig.~\ref{fig:thickness}, the gray band marks the region spanned by $\T_\text{min}$ and $\T_\text{max}$, while the blue band and line show the generalized mean of the participant densities for different values of the parameter $p$.
We see that decreasing $p$ pulls the generalized mean towards the minimum of $\T_A$ and $\T_B$ while increasing $p$ pushes it to the maximum, thus, the generalized mean ansatz parametrizes asymmetric entropy deposition, or in the parlance of color glass condensate theory, the intensity of saturation effects on local gluon production.

\fig{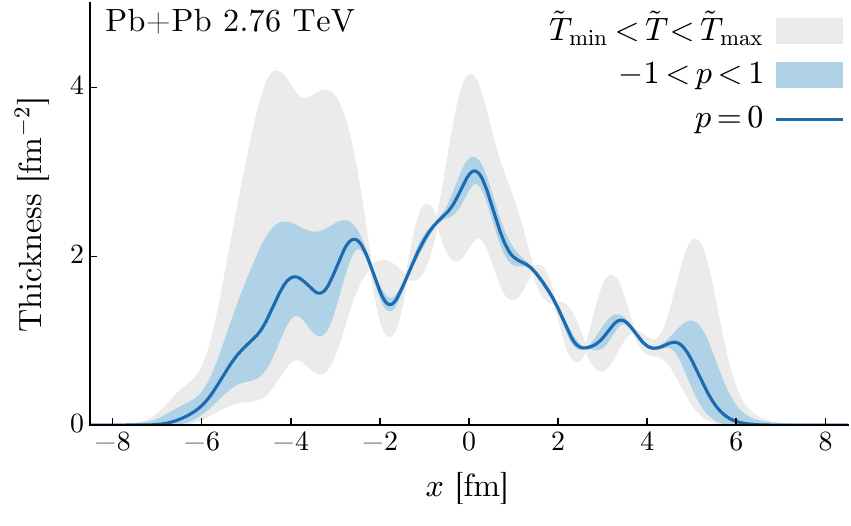}{
    Cross section of the participant nucleon density in a mid-central Pb+Pb collision at $\sqrts=2.76$ TeV as a function of the transverse coordinate $x$ parallel to impact parameter $b$.
  The gray band indicates the region bounded by the minimum and maximum values of the local participant thickness functions $\T_A$ and $\T_B$, while the blue band indicates the region spanned by the generalized mean of $\T_A$ and $\T_B$ with parameter $-1<p<1$.
  The solid blue line shows an example of a discrete mapping specified by a generalized mean with $p=0$.
}

\fig[b]{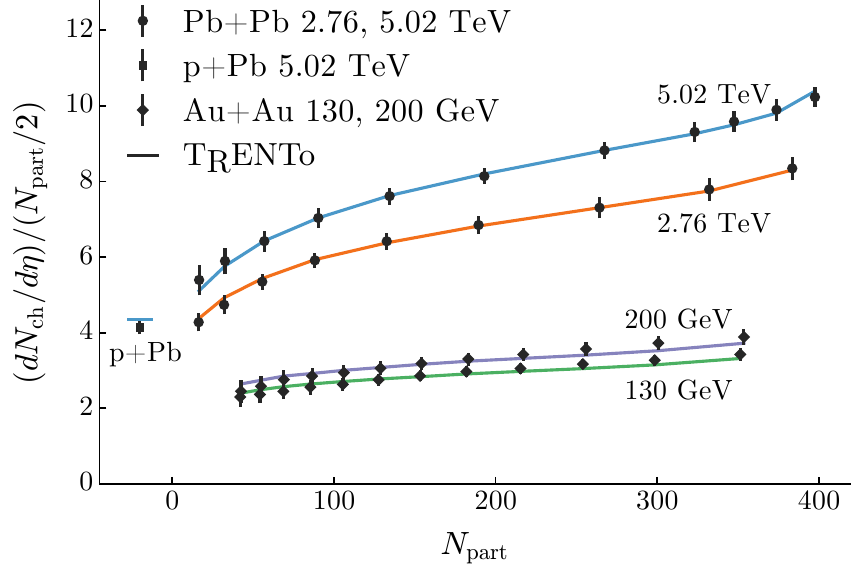}{
    Average charged particle density per participant pair $(d\nch/d\eta)/(\npart/2)$ at midrapidity as a function of participant number for Pb+Pb, p+Pb, and Au+Au systems at various collision energies.
    Lines are \protect\trento\ calculations with generalized mean parameter $p=0$, and symbols are experimental data from PHENIX \cite{Adare:2015bua} and ALICE \cite{Aamodt:2010cz, Adam:2015ptt}.
    The average minimum bias participant number for p+Pb is shifted for clarity.
}

These local modifications naturally become manifest in global quantities such as integrated particle yields.
When two heavy ions collide at fixed impact parameter $b$, their nuclear densities are shifted by a common offset $T(x\pm b/2,y)$ which increases the average asymmetry of local participant matter.
This asymmetry grows with increasing impact parameter and is highly correlated with collision centrality.
By varying the generalized mean parameter $p$, the \trento\ model directly modulates the attenuation of entropy deposition in peripheral collisions and provides a parametric handle on the centrality dependence of charged particle production---similar to the role of the binary collision fraction $\alpha$ in the two-component Glauber model.

Figure~\ref{fig:nch_per_npart} plots the charged particle density per participant pair at midrapidity as a function of participant number using model calculations from \trento\ and experimental data from PHENIX \cite{Adare:2015bua} and ALICE \cite{Aamodt:2010cz, Adam:2015ptt}.
The model curves are calculated assuming that charged particle multiplicity is proportional to total initial entropy \cite{Song:2008si}, where the proportionality constant varies with beam energy but is constant for all collision systems at the same energy.
We set the entropy deposition parameter $p=0$, which was previously shown to provide a good description of proton-proton, proton-lead, and lead-lead multiplicity distributions as well as lead-lead eccentricity harmonics at LHC energies \cite{Moreland:2014oya}.
However, this value and the other parameters used in Fig.~\ref{fig:nch_per_npart} have not yet been systematically optimized---they are for demonstration purposes only.
While $p$ could depend on energy, we see in the figure that $p=0$ provides a good description of the data at all beam energies and self-consistently describes proton-lead and lead-lead multiplicities at the same collision energy.

\fig*{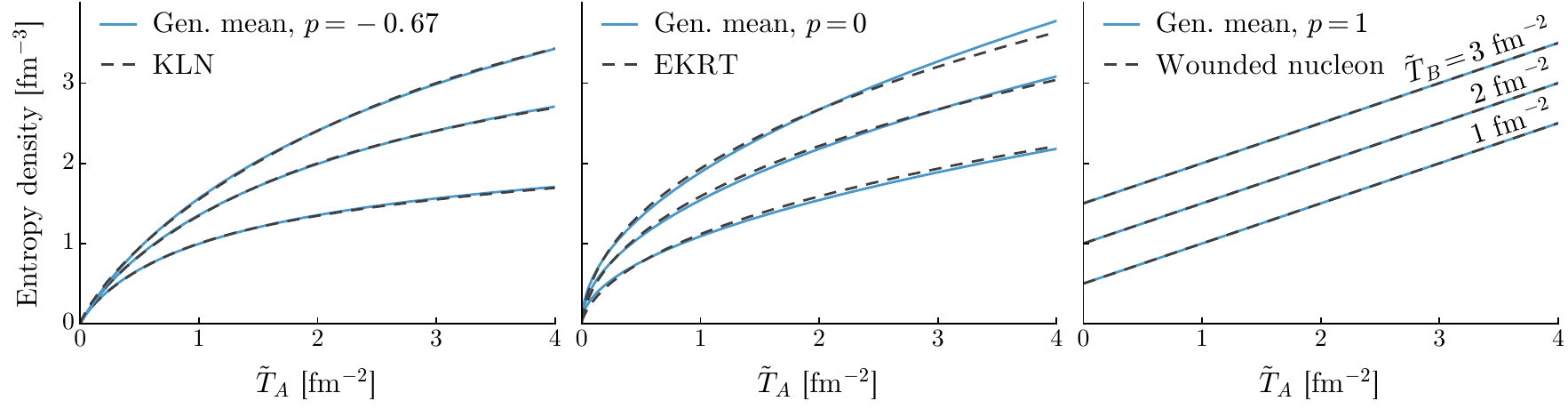}{
  Profiles of the initial thermal distribution predicted by the KLN (left), EKRT (middle), and wounded nucleon (right) models (dashed black lines) compared to a generalized mean with different values of the parameter $p$ (solid blue lines).
  Staggered lines show different slices of the initial entropy density $dS/(d^2r_\perp dy)$ as a function of the participant nucleon density $\T_A$ for several values of $\T_B = 1, 2, 3$ [fm$^{-2}$].
  The EKRT mapping is shown with model parameters $K=0.64$ and $\beta=0.8$ \cite{Niemi:2015qia}.
  Entropy normalization is arbitrary.
}

Note that, while the generalized mean parametrizes entropy deposition in asymmetric regions of the collision ($\T_A \neq \T_B$), it asserts a particular scaling in symmetric regions, namely
\begin{equation}
  f(\alpha \T, \alpha \T) = \alpha \T,
  \label{eq:homogenous}
\end{equation}
for a constant $\alpha$.
This property, known as scale invariance or homogeneity, is difficult to empirically prove or disprove, but multiple experimental observations indicate that it holds to very good approximation.
For example, it was demonstrated that collisions of highly deformed uranium nuclei exhibit elliptic flow patterns which are incompatible with a scale-violating binary collision term postulated by the two-component Glauber ansatz \cite{Goldschmidt:2015kpa, Pandit:2013uiv, Wang:2014qxa}.
Measurements of central copper-copper, gold-gold, and uranium-uranium particle production at RHIC also exhibit approximate participant scaling \cite{Adare:2015bua}.
Moreover, the scale invariant constraint serves as a reasonable approximation for a number of calculations of the mapping $f$ in Eq.~\eqref{eq:mapping} derived from CGC effective field theory, as we show momentarily.
At present, we thus assert scale invariance as a simplifying postulate, although relaxing this constraint may further reduce bias and could be considered in future work.

\subsection{Reproducing existing I.C.\ models}

The aforementioned procedure defines the \trento\ initial condition model proposed in Ref.~\cite{Moreland:2014oya}.
The model is constructed to achieve maximal flexibility using a minimal number of parameters and can mimic a wide range of existing initial condition models.
To demonstrate the efficacy of the generalized mean ansatz, Eq.~\eqref{eq:genmean}, we now show that the mapping can reproduce different theory calculations using suitable values of the parameter $p$.

Perhaps the simplest and oldest model of heavy-ion initial conditions is the so called participant or wounded nucleon model, which deposits entropy for each nucleon that engages in one or more inelastic collisions \cite{Bialas:1976ed}.
In its Monte Carlo formulation \cite{Shor:1988vk, Wang:1991hta, Alver:2008aq, Broniowski:2007nz}, the wounded nucleon model may be expressed in terms of participant thickness functions, Eq.~\eqref{eq:participant}, as
\begin{equation}
  s \propto \T_A + \T_B.
  \label{eq:wn}
\end{equation}
Comparing to Eq.~\eqref{eq:means}, we see that the wounded nucleon model is equivalent to the generalized mean ansatz with $p=1$.

More sophisticated calculations of the mapping $f$ in Eq.~\eqref{eq:mapping} can be derived from color glass condensate effective field theory.
A common implementation of a CGC based saturation picture is the KLN model \cite{Kharzeev:2001yq, Kharzeev:2002ei, Kharzeev:2004if}, in which entropy deposition at the QGP thermalization time can be determined from the produced gluon density, $s \propto N_g$, where
\begin{equation}
  \frac{dN_g}{dy\,d^2r_\perp} \sim \Qs{min}^2 \biggl[
    2 + \log \biggl(\frac{\Qs{max}^2}{\Qs{min}^2} \biggr)
  \biggr],
  \label{eq:kln}
\end{equation}
and $\Qs{max}$ and $\Qs{min}$ denote the larger and smaller values of the two saturation scales in opposite nuclei at any fixed position in the transverse plane \cite{Drescher:2006ca}.
In the original formulation of the KLN model, the two saturation scales are proportional to the local participant nucleon density in each nucleus, $Q^2_{s,A} \propto \T_A$, and the gluon density can be recast as
\begin{equation}
  s \sim \T_\text{min} \bigl[ 2 + \log(\T_\text{max}/\T_\text{min}) \bigr]
\end{equation}
to put it in a form which can be directly compared with the wounded nucleon model.

Another saturation model which has attracted recent interest after it successfully described an extensive list of experimental particle multiplicity and flow observables \cite{Niemi:2015qia, Paatelainen:2013eea} is the EKRT model, which combines collinearly factorized pQCD minijet production with a simple conjecture for gluon saturation \cite{Eskola:1999fc, Eskola:2001bf}.
The energy density predicted by the model after a pre-thermal Bjorken free streaming stage is given by
\begin{equation}
  e(\tau_0, x, y) \sim \frac{K_\text{sat}}{\pi} p_\text{sat}^3(K_\text{sat}, \beta; T_A, T_B),
  \label{eq:ekrt_energy}
\end{equation}
where the saturation momentum $p_\text{sat}$ depends on nuclear thickness functions $T_A$ and $T_B$, as well as phenomenological model parameters $K_\text{sat}$ and $\beta$.
Calculating the saturation momentum in the EKRT formalism is computationally intensive, and hence---in its Monte Carlo implementation---the model parametrizes the saturation momentum $p_\text{sat}$ to facilitate efficient event sampling \cite{Niemi:2015qia}.
The energy density in Eq.~\eqref{eq:ekrt_energy} can then be recast as an entropy density using the thermodynamic relation ${s \sim e^{3/4}}$ to compare it with the previous models.

Note that Eq.~\eqref{eq:ekrt_energy} is expressed as a function of nuclear thickness $T$ which includes contributions from \emph{all} nucleons in the nucleus, as opposed to the participant thickness $\T$.
In order to express initial condition mappings as functions of a common variable one could, e.g.\ relate $\T$ and $T$ using an analytic wounded nucleon model.
The effect of this substitution on the EKRT model is small, as the mapping deposits zero entropy if nucleons are non-overlapping, effectively removing them from the participant thickness function.
We thus replace $T$ with $\T$ in the EKRT model and note that similar results are obtained by recasting the wounded nucleon, KLN, and \trento\ models as functions of $T$ using standard Glauber relations.

Figure~\ref{fig:cgc_compare} shows one-dimensional slices of the entropy deposition mapping predicted by the KLN, EKRT, and wounded nucleon models for typical values of the participant nucleon density sampled in Pb+Pb collisions at $\sqrts=2.76$~TeV.
The vertically staggered lines in each panel show the change in deposited entropy density as a function of $\T_A$ for several constant values of $\T_B$, where the dashed lines are the entropy density calculated using the various models and the solid lines show the generalized mean ansatz tuned to fit each model.
The figure illustrates that the ansatz reproduces different initial condition calculations and quantifies differences among them in terms of the generalized mean parameter $p$.
The KLN model, for example, is well-described by $p\sim-0.67$, the EKRT model corresponds to $p \sim 0$, and the wounded nucleon model is precisely $p=1$.
Smaller, more negative values of $p$ pull the generalized mean toward a minimum function and hence correspond to models with more extreme gluon saturation effects.

\fig{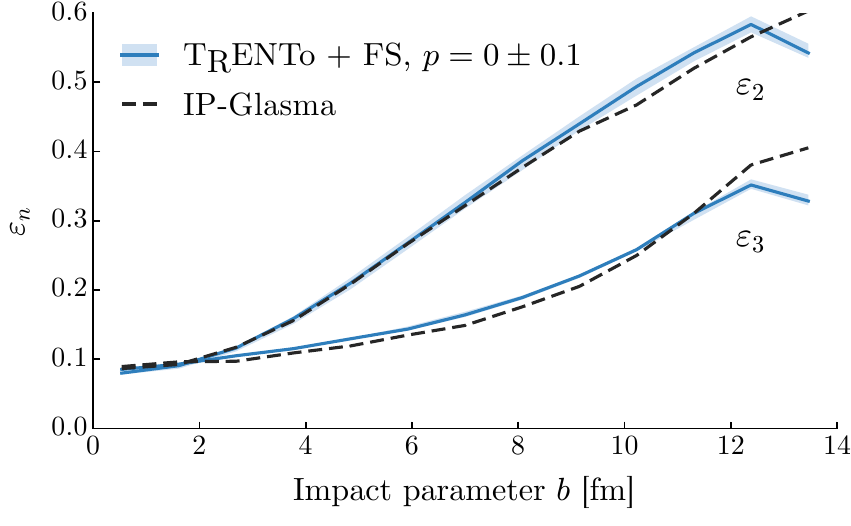}{
    Eccentricity harmonics $\varepsilon_2$ and $\varepsilon_3$ as a function of impact parameter $b$ for Pb+Pb collisions at ${\sqrts=2.76}$~TeV calculated from IP-Glasma and \protect\trento\ initial conditions.
  IP-Glasma events are evaluated after $\tau=0.4$~fm/$c$ classical Yang-Mills evolution \cite{Schenke:2012wb}; \protect\trento\ events after $\tau=0.4$~fm/$c$ free streaming \cite{Liu:2015nwa, Broniowski:2008qk} and using parameters $p = 0 \pm 0.1$, $k = 1.6$, and nucleon width $w=0.4$~fm to match IP-Glasma \cite{Schenke:2013dpa}.
}

The three models considered in Fig.~\ref{fig:cgc_compare} are by no means an exhaustive list of proposed initial condition models, see e.g.\ Refs.~\cite{Eremin:2003qn, Broniowski:2007nz, Pierog:2013ria, Drescher:2000ec, Chatterjee:2015aja, Zhang:1999bd}.
Notably absent, for instance, is the highly successful IP-Glasma model which combines IP-Sat CGC initial conditions with classical Yang-Mills dynamics to describe the full pre-equilibrium evolution of produced glasma fields \cite{Schenke:2012wb, Schenke:2012fw, Gale:2012rq}.
The IP-Glasma model lacks a simple analytic form for initial energy (or entropy) deposition at the QGP thermalization time and so cannot be directly compared to the generalized mean ansatz.
In lieu of such a comparison, we examined the geometric properties of IP-Glasma and \trento\ through their eccentricity harmonics $\varepsilon_n$.

We generated a large number of \trento\ events using entropy deposition parameter $p=0$, Gaussian nucleon width $w=0.4$~fm, and fluctuation parameter $k=1.6$, which were previously shown to reproduce the ratio of ellipticity and triangularity in IP-Glasma \cite{Moreland:2014oya}.
We then free streamed \cite{Liu:2015nwa, Broniowski:2008qk} the events for $\tau=0.4$~fm/$c$ to mimic the weakly coupled pre-equilibrium dynamics of IP-Glasma and match the evolution time of both models.
Finally, we calculated the eccentricity harmonics $\varepsilon_2$ and $\varepsilon_3$ weighted by energy density $e(x, y)$ according to the definition
\begin{equation}
    \varepsilon_n e^{i n \phi} = -\frac{\int dx\, dy\, r^n e^{i n \phi} e(x,y)}{\int dx\, dy\, e(x,y)},
\end{equation}
where the energy density is the time-time component of the stress-energy tensor after the free streaming phase, $T^{00}$.
The resulting eccentricities, pictured in Fig.~\ref{fig:ipglasma}, are in good agreement for all but the most peripheral collisions, where sub-nucleonic structure becomes important.
This similarity suggests that \trento\ with $p \sim 0$ can effectively reproduce the scaling behavior of IP-Glasma, although a more detailed comparison would be necessary to establish the strength of correspondence illustrated in Fig.~\ref{fig:cgc_compare}.

Additionally, a participant quark model has been proposed to describe the multiplicity and transverse-energy distributions of a variety of collision systems without a binary collision term \cite{Adler:2013aqf, Adare:2015bua}.
The model can be recast using an analytic Glauber formalism to construct an effective entropy deposition mapping in the form of Eq.~\eqref{eq:mapping}.
However, the resulting mapping cannot be encapsulated by a single value of the parameter $p$, so we do not attempt to support or exclude the participant quark model in the present analysis.

\section{Parameter estimation}

With the full evolution model in hand, a number of important model parameters---related to both initial-state entropy deposition and the QGP medium---remain undetermined.
These parameters typically correlate among each other and affect multiple observables, hence, if we wish to describe a wide variety of experimental observables, the only option is a simultaneous fit to all parameters.
However, it is not feasible to do this directly, since simulating observables at even a single set of parameter values requires thousands of individual events and significant computation time.

To overcome this limitation, we employ a Bayesian method for parameter estimation with computationally expensive models \cite{OHagan:2006ba,Higdon:2008cmc,Higdon:2014tva,Wesolowski:2015fqa}.
Briefly, the model is evaluated at a relatively small \order 2 number of parameter points, the output is interpolated by a Gaussian process emulator, and the emulator is used to systematically explore the parameter space with Markov chain Monte Carlo methods.
This section summarizes the methodology; see Ref.~\cite{Bernhard:2015hxa} for a complete treatment.

\subsection{Model parameters and observables}

We choose a set of nine model parameters for estimation.
Four control the parametric initial state:
\begin{enumerate}[itemsep=0pt]
  \item the overall normalization factor,
  \item entropy deposition parameter $p$ from the generalized mean ansatz Eq.~\eqref{eq:genmean},
  \item gamma shape parameter $k$, which sets nucleon multiplicity fluctuations in Eq.~\eqref{eq:participant}, and
  \item Gaussian nucleon width $w$ from Eq.~\eqref{eq:Tp}, which determines initial-state granularity;
\end{enumerate}
the remaining five are related to the QGP medium:
\begin{enumerate}[itemsep=0pt]
  \item[5--7.] the three parameters ($\eta/s$ hrg, min, and slope) in Eq.~\eqref{eq:etas} that set the temperature dependence of the specific shear viscosity,
  \setcounter{enumi}{7}
  \item normalization prefactor for the temperature dependence of bulk viscosity Eq.~\eqref{eq:zetas}, and
  \item particlization temperature $T_\text{switch}$.
\end{enumerate}
This parameter set will enable simultaneous characterization of the initial state and medium, including any correlations.
Table~\ref{tab:design} summarizes the parameters and their corresponding ranges, which are intentionally wide to ensure that the optimal values are bracketed.

\begin{table}
  \caption{
    \label{tab:design}
    Input parameter ranges for the initial condition and hydrodynamic models.
  }
  \begin{ruledtabular}
  \begin{tabular}{lll}
    Parameter         & Description                        & Range           \\
    \paddedhline
    Norm              & Overall normalization              & 100--250        \\
    $p$               & Entropy deposition parameter       & $-1$ to $+1$    \\
    $k$               & Multiplicity fluct.\ shape         & 0.8--2.2        \\
    $w$               & Gaussian nucleon width             & 0.4--1.0 fm     \\
    $\eta/s$ hrg      & Const.\ shear viscosity, $T < T_c$ & 0.3--1.0        \\
    $\eta/s$ min      & Shear viscosity at $T_c$           & 0--0.3          \\
    $\eta/s$ slope    & Slope above $T_c$                  & 0--2 GeV$^{-1}$ \\
    $\zeta/s$ norm    & Prefactor for $(\zeta/s)(T)$       & 0--2            \\
    $T_\text{switch}$ & Particlization temperature         & 135--165 MeV    \\
  \end{tabular}
  \end{ruledtabular}
\end{table}

\begin{table*}
  \caption{
    \label{tab:observables}
    Experimental data to be compared with model calculations.
  }
  \begin{ruledtabular}
  \begin{tabular}{lcccc}
    Observable & Particle species & Kinematic cuts & Centrality classes & Ref. \\
    \paddedhline
    Yields $dN/dy$                       & $\pi^\pm$, $K^\pm$, $p\bar p$ &
    $|y| < 0.5$ & 0--5, 5--10, 10--20, \ldots, 60--70 & \cite{Abelev:2013vea} \\
    \noalign{\smallskip}
    Mean transverse momentum $\avg{p_T}$ & $\pi^\pm$, $K^\pm$, $p\bar p$ &
    $|y| < 0.5$ & 0--5, 5--10, 10--20, \ldots, 60--70 & \cite{Abelev:2013vea} \\
    \noalign{\smallskip}
    Two-particle flow cumulants $\vnk n 2$ & \multirow{2}{*}{all charged} &
    $|\eta| < 1$ & 0--5, 5--10, 10--20, \ldots, 40--50 &
    \multirow{2}{*}{\cite{ALICE:2011ab}} \\
    $n = 2$, 3, 4 & & $0.2 < p_T < 5.0$ GeV & $n = 2$ only: 50--60, 60--70 & \\
  \end{tabular}
  \end{ruledtabular}
\end{table*}

Having designated the model parameters and ranges, we generated a 300 point maximin\footnote{A ``maximin'' design \emph{maximizes} the \emph{minimum} distance between points, thereby reducing large gaps and tight clusters.} Latin hypercube design \cite{Morris:1995lh} in the nine-dimensional parameter space and executed \order 4 minimum-bias Pb+Pb events at each of the 300 points.
Each event consists of a single ``bumpy'' (i.e.\ Monte Carlo sampled) initial condition and hydro simulation followed by multiple samples of the freeze-out hypersurface.
The number of samples is roughly inversely proportional to the event's particle multiplicity so that total particle production is constant across all events---typically ${\sim}$5 samples for central events and up to 100 for peripheral events.
This strategy leads to consistent statistical uncertainties across all parameter points and centrality classes.

Parameter estimation relies on observables that are sensitive to varying the model inputs.
For example, bulk viscosity suppresses radial expansion, so a meaningful estimate of the $(\zeta/s)(T)$ normalization parameter requires some measure of radial flow such as the mean transverse momentum.
Indeed, previous work has shown that finite bulk viscosity is necessary to simultaneously fit both mean transverse momentum and anisotropic flow \cite{Ryu:2015vwa}.

For the present study we compare to the centrality dependence of identified particle yields $dN/dy$ and mean transverse momenta $\avg{p_T}$ for charged pions, kaons, and protons as well as two-particle anisotropic flow coefficients $\vnk n 2$ for $n = 2$, 3, 4.
Table~\ref{tab:observables} summarizes the observables including kinematic cuts, centrality classes, and experimental data, which are all from the ALICE experiment, Pb+Pb collisions at $\sqrts = 2.76$ TeV \cite{Abelev:2013vea,ALICE:2011ab}.
These observables characterize the lowest-order moments of the transverse momentum and flow distributions;
including higher-order quantities such as mean-square momenta $\avg{p_T^2}$ \cite{Heinz:2015arc} and four-particle cumulants $\vnk n 4$ \cite{Aamodt:2010pa} could enable a more precise fit.

When computing simulated observables, we strive to replicate experimental methods as closely as possible.
We selected the same centrality classes as the corresponding experimental data by sorting each design point's minimum-bias events by charged-particle multiplicity $d\nch/d\eta$ at midrapidity ($|\eta| < 0.5$) and dividing the events into the desired percentile bins.
We computed identified $dN/dy$ and $\avg{p_T}$ by simple counting and averaging of the desired species at midrapidity ($|y| < 0.5$); no additional steps are necessary since the experimental data are corrected and extrapolated to zero $p_T$ \cite{Abelev:2013vea}.
Finally, we calculated flow coefficients for charged particles within the kinematic range of the ALICE detector using the direct $Q$-cumulant method \cite{Bilandzic:2010jr}.

The top row of Fig.~\ref{fig:observables_samples} (located later in Sec.~\ref{sec:results}) shows the final observables for each of the 300 design points;
their large spreads arise from the wide input parameter ranges.

\subsection{Gaussian process emulators}

\newcommand{\x}{\mathbf x}
\newcommand{\y}{\mathbf y}
\newcommand{\N}{\mathcal N}
\newcommand{\muvec}{\boldsymbol\mu}
\newcommand{\tran}{^\intercal}

Central to the parameter estimation method is a statistical surrogate model that interpolates the model input parameter space and provides fast predictions of the output observables at arbitrary inputs.
We use Gaussian process emulators \cite{Rasmussen:2006gp} as flexible, non-parametric interpolators.
Essentially, this amounts to assuming that the model follows a multivariate normal distribution with mean and covariance functions determined by conditioning on actual model calculations.

The full evolution model takes vectors $\x$ of $n = 9$ inputs and produces a number of outputs (each centrality bin of each observable is an output).
For the moment consider only a single output, e.g.\ pion $dN/dy$ in \mbox{20--30\%} centrality (the specific observable does not matter), and call it $y$.
We have already evaluated the model at $m = 300$ design points, i.e.\ an $m \times n$ design matrix $X = \{\x_1, \ldots, \x_m\}$, and obtained the corresponding $m$ outputs $\y = \{y_1, \ldots, y_m\}$.
Now, we assume that the model is a Gaussian process with some covariance function $\sigma$ and condition it on the training data $(X, \y)$, yielding predictions for the outputs $\y_*$ at some other points $X_*$ within the design range.
The predictive distribution for $\y_*$ is the multivariate normal distribution
\begin{equation}
  \begin{aligned}
    \y_* &\sim \N(\muvec, \Sigma), \\
    \muvec &= \sigma(X_*, X)\sigma(X, X)^{-1}\y, \\
    \Sigma &= \sigma(X_*,X_*) - \sigma(X_*,X)\sigma(X,X)^{-1}\sigma(X,X_*),
  \end{aligned}
\end{equation}
where $\muvec$ is the mean vector and $\Sigma$ the covariance matrix, and the notation $\sigma(\cdot, \cdot)$ indicates a matrix from applying the covariance function to each pair of inputs, e.g.
\begin{equation}
  \sigma(X, X) =
  \begin{pmatrix}
    \sigma(\x_1, \x_1) & \cdots & \sigma(\x_1, \x_m) \\
    \vdots & \ddots & \vdots \\
    \sigma(\x_m, \x_1) & \cdots & \sigma(\x_m, \x_m) \\
  \end{pmatrix}.
\end{equation}
Thus, we obtain both the mean predicted output and corresponding uncertainty at any desired input point.
Generally, the uncertainty is small near explicitly calculated points and wide in gaps, reflecting the true state of knowledge of the interpolation.

The covariance function $\sigma$ quantifies the correlation between pairs of input points.
We use a typical Gaussian function
\begin{equation}
  \sigma(\x, \x') = \sigma_\text{GP}^2 \exp\Biggl[ -\sum_{k=1}^n \frac{(x_k - x'_k)^2}{2\ell_k^2} \Biggr] + \sigma_n^2\delta_{\x\x'},
\end{equation}
which yields smoothly-varying processes with continuous derivatives, making it a common choice for well-behaved models.
This form has several variable \emph{hyperparameters}:
the overall variance of the Gaussian process $\sigma_\text{GP}^2$,
the correlation lengths for each input parameter $\ell_k$,
and the noise variance $\sigma_n^2$ which allows for statistical error.
These hyperparameters may be estimated from the training data by numerically maximizing the likelihood function
\begin{equation}
  \log P = -\frac{1}{2} \y\tran \Sigma^{-1} \y - \frac{1}{2} \log |\Sigma| - \frac{m}{2} \log 2\pi,
\end{equation}
with $\Sigma = \sigma(X, X)$, i.e.\ the covariance function applied to the inputs.
This expression consists of a least-squares fit to the data (first term), a complexity penalty to prevent overfitting (second term), and a normalization constant (third term).

\fig*{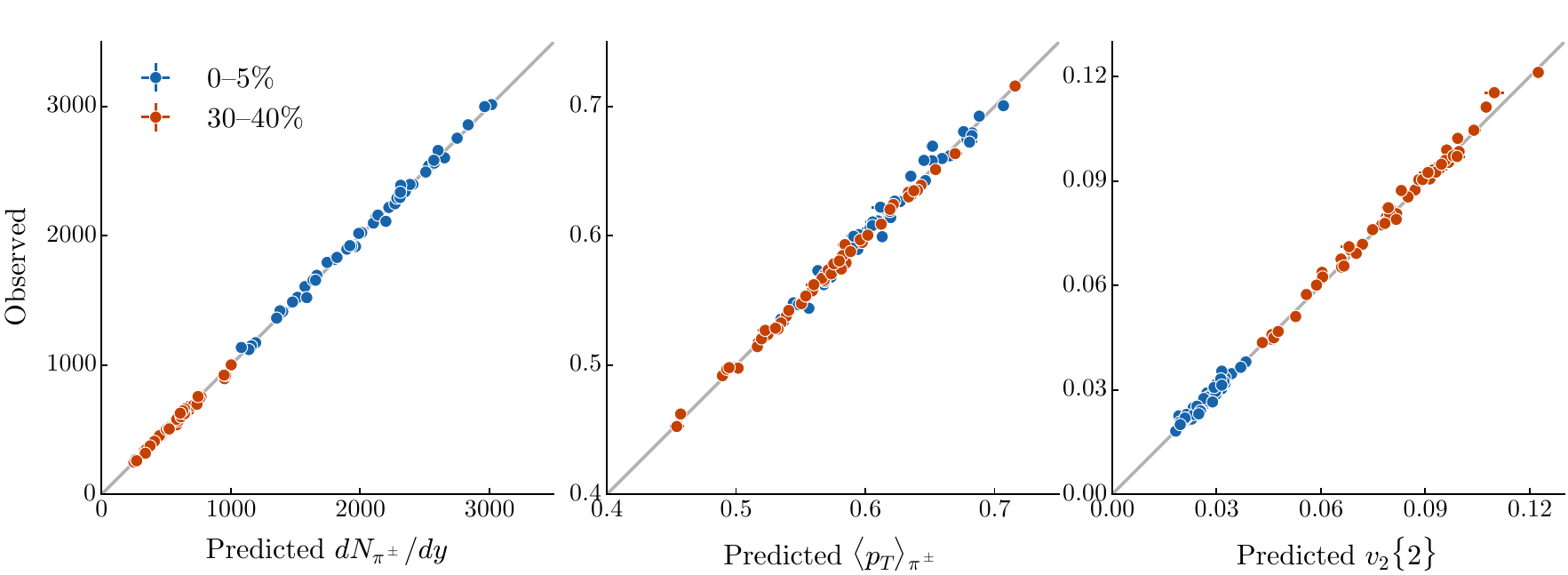}{
  Validation of Gaussian process emulator predictions.
  Each panel shows predictions compared to explicit model calculations at the 50 validation design points.
  The horizontal location and error bar of each point indicates the predicted value and uncertainty,
  vertical indicates the explicitly calculated value and statistical uncertainty,
  and the diagonal gray line represents perfect agreement.
  Left: charged pion yields $dN_{\pi^\pm}/dy$,
  middle: mean pion transverse momenta $\avg{p_T}_{\pi^\pm}$,
  right: flow cumulant $\vnk 2 2$;
  each in centrality bins 0--5\% (blue) and 30--40\% (orange).
}

To this point we have considered only a single output.
Gaussian processes are fundamentally scalar functions, but the model produces many outputs, all of which must be emulated.
This is readily solved by transforming the output data into orthogonal and uncorrelated linear combinations called principal components, then emulating each component with an individual Gaussian process.

Let $p$ be the number of model outputs, that is, given an $m \times n$ design matrix $X$, the model produces an $m \times p$ output matrix $Y$.
The principal components $Z$ are then computed by the linear transformation
\begin{equation}
  Z = \sqrt m \, Y U
\end{equation}
where $U$ are the eigenvectors of the sample covariance matrix $Y\tran Y$.
The Gaussian processes predict principal components $Z_*$ at input points $X_*$ which are then transformed back to physical space as
\begin{equation}
  Y_* = \frac{1}{\sqrt m} Z_* U\tran.
\end{equation}

Often, the $p$ model outputs are strongly correlated and so a much smaller number of principal components $q~\ll~p$ account for most of the model's variance.
Thus one can use only $q$ components, reducing a high-dimensional output space to a few one-dimensional problems with negligible loss of information.
We use $q = 8$ principal components, retaining over 99.5\% of the variance from the original $p = 68$ outputs.

To validate the performance of the emulators, we generated an independent 50 point Latin hypercube design from the original design space, evaluated the full model at each validation point, and compared the explicit model calculations to emulator predictions.
Figure~\ref{fig:validation} confirms that the emulators faithfully predict true model calculations.
The predictions need not agree perfectly at every point; ideally the residuals would be normally distributed with mean zero and variance predicted by the Gaussian processes.

\subsection{Bayesian calibration}

\fig*{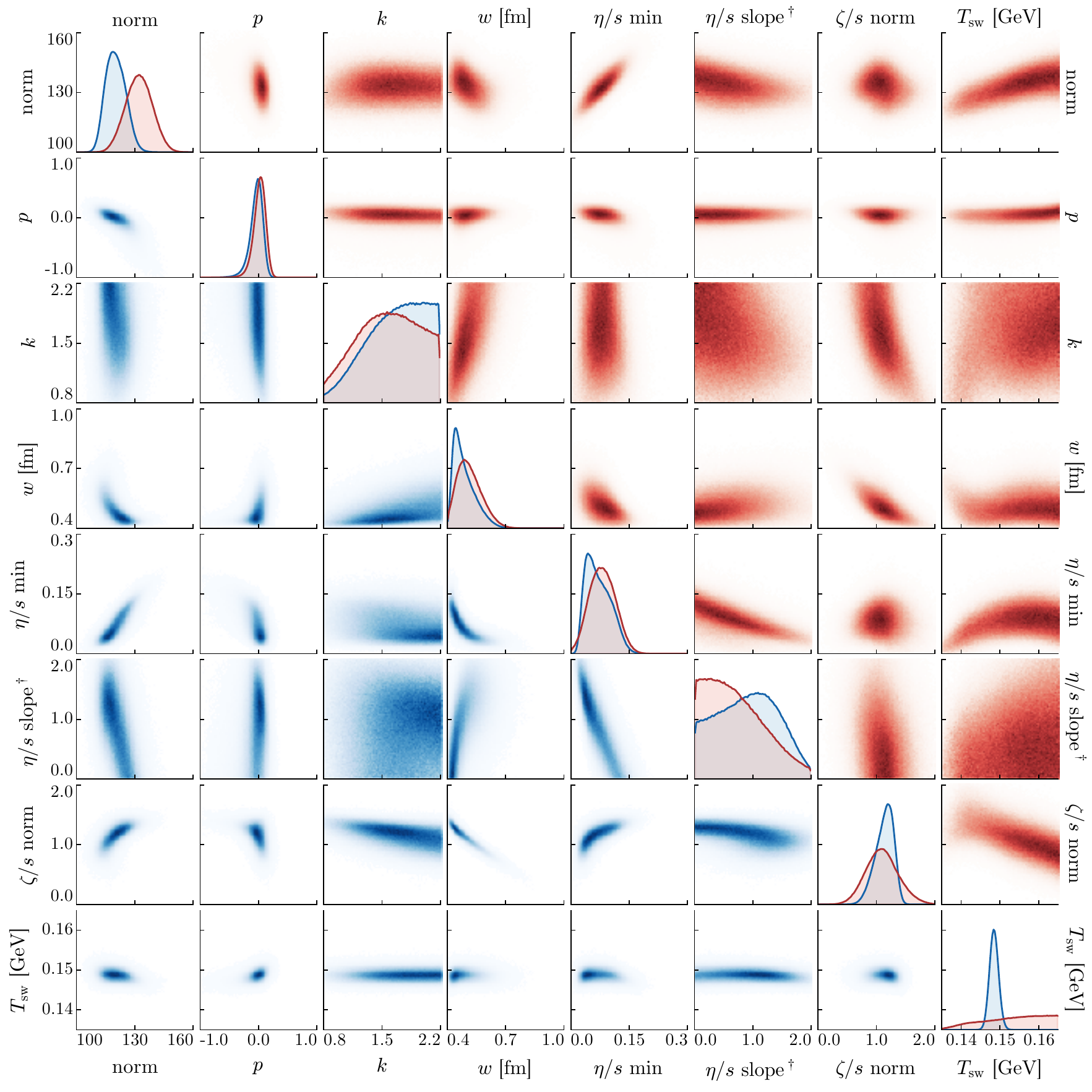}{
  Posterior distributions for the model parameters from calibrating to identified particles yields (blue, lower triangle) and charged particles yields (red, upper triangle).
  The diagonal has marginal distributions for each parameter, while the off-diagonal contains joint distributions showing correlations among pairs of parameters.
  $^\dagger$The units for $\eta/s$ slope are [GeV$^{-1}$].
}

\fig*{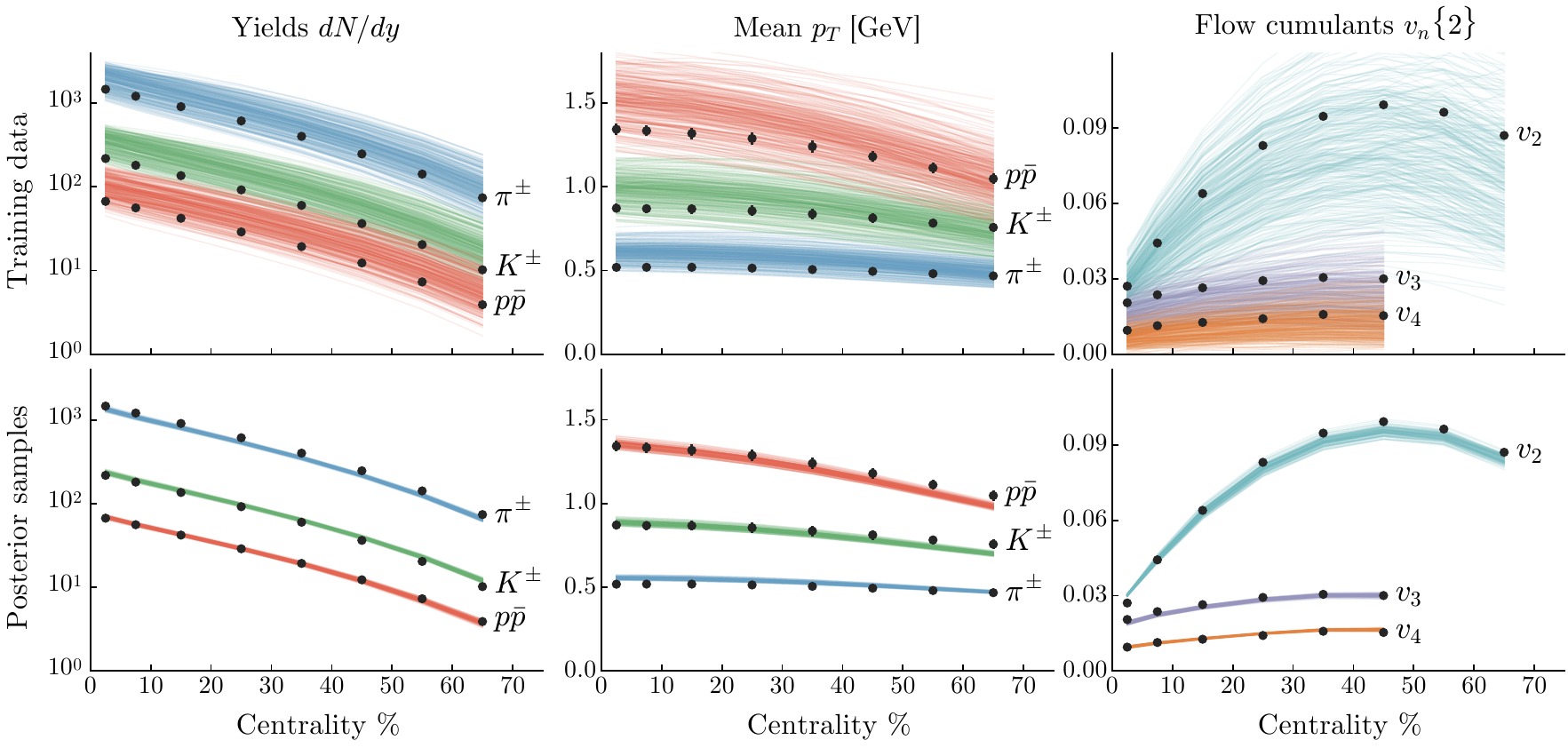}{
  Simulated observables compared to experimental data from the ALICE experiment \cite{Abelev:2013vea,ALICE:2011ab}.
  Top row: explicit model calculations for each of the 300 design points,
  bottom: emulator predictions of 100 random samples drawn from the posterior distribution.
  Left column: identified particle yields $dN/dy$,
  middle: mean transverse momenta $\avg{p_T}$,
  right: flow cumulants $v_n\{2\}$.
}

\newcommand{\z}{\mathbf z}
\newcommand{\st}{_\star}
\newcommand{\ex}{_\text{exp}}

The final step in the parameter estimation method is to calibrate the model parameters to optimally reproduce experimental observables, thereby extracting probability distributions for the true values of the parameters.
According to Bayes' theorem, the probability for the true parameters $\x\st$ is
\begin{equation}
  P(\x\st|X,Y,\y\ex) \propto P(X,Y,\y\ex|\x\st) P(\x\st).
  \label{eq:bayes}
\end{equation}
The left-hand side is the \emph{posterior}: the probability of $\x\st$ given the design $X$, computed observables $Y$, and experimental data $\y\ex$.
On the right-hand side, $P(\x\st)$ is the \emph{prior} probability---encapsulating initial knowledge of $\x\st$---and $P(X,Y,\y\ex|\x\st)$ is the likelihood: the probability of observing $(X, Y, \y\ex)$ given a proposal $\x\st$.

The likelihood may be quickly computed using the principal component Gaussian process emulators constructed in the previous subsection:
\begin{align}
  P &= P(X,Y,\y\ex|\x\st) \nonumber \\
    &= P(X,Z,\z\ex|\x\st) \nonumber \\
    &\propto\exp\biggl\{
      -\frac{1}{2} (\z\st - \z\ex)\tran \Sigma_z^{-1} (\z\st - \z\ex)
    \biggr\},
  \label{eq:likelihood}
\end{align}
where $\z\st = \z\st(\x\st)$ are the principal components predicted by the emulators, $\z\ex$ is the principal component transform of the experimental data $\y\ex$, and $\Sigma_z$ is the covariance (uncertainty) matrix.
As in previous work \cite{Novak:2013bqa,Bernhard:2015hxa}, we assume a constant fractional uncertainty on the principal components, so that the covariance matrix is
\begin{equation}
  \Sigma_z = \text{diag}(\sigma^2_z\,\z\ex),
  \label{eq:uncertainty}
\end{equation}
with $\sigma_z = 0.10$ in the present study.
This is a simple ansatz intended to conservatively account for the various sources of uncertainty in the experimental data, model calculations, and emulator predictions.
It certainly limits the meaning of quantitative uncertainties in the final estimated parameters and is an obvious target for improvement in future studies.

We place a uniform prior on the model parameters, i.e.\ the prior is constant within the design range and zero outside.
Combined with the likelihood \eqref{eq:likelihood} and Bayes' theorem \eqref{eq:bayes}, we can easily evaluate the posterior probability at any point in parameter space.

\enlargethispage{.5\baselineskip}

Posterior distributions are typically constructed using Markov chain Monte Carlo (MCMC) methods.
MCMC algorithms generate random walks through parameter space by accepting or rejecting proposal points based on the posterior probability; after many steps the chain converges to the desired posterior.

We use the affine-invariant ensemble sampler \cite{Goodman:2010en,FM:2013mc}, an efficient MCMC algorithm that uses a large ensemble of interdependent walkers.
We first run \order 6 steps to allow the chain to equilibrate, discard these ``burn-in'' samples, then generate \order 7 posterior samples.

\vfill

\section{\label{sec:results}Results}

The primary result of this study is the posterior distribution for the model parameters, Fig.~\ref{fig:posterior}.
In fact, this figure contains two posterior distributions:
one from calibrating to identified particle yields $dN/dy$ (blue, lower triangle),
and the other from calibrating to charged particle yields $d\nch/d\eta$ (red, upper triangle).
We performed the alternate calibration to charged particles because the model could not simultaneously describe all identified particle yields for \emph{any} parameter values, as will be demonstrated shortly.

In Fig.~\ref{fig:posterior}, the diagonal plots are marginal distributions for each model parameter (all other parameters integrated out) from the calibrations to identified (blue) and charged (red) particles, while the off-diagonals are joint distributions showing correlations among pairs of parameters from the calibrations to identified (blue, lower triangle) and charged (red, upper triangle) particles.
Operationally, these are all histograms of MCMC samples.

We discuss the posterior distributions in detail in the following subsections.
First, let us introduce several ancillary results.

Table~\ref{tab:posterior} contains quantitative estimates of each parameter extracted from the posterior distributions.
The reported values are the medians of each parameter's distribution, and the uncertainties are highest-posterior density\footnote{The highest-posterior density credible interval is the smallest range containing the desired fraction of the distribution.} 90\% credible intervals.
Note that some estimates are influenced by limited prior ranges, e.g.\ the lower bound of the nucleon width $w$.

Figure~\ref{fig:observables_samples} compares simulated observables (see Table~\ref{tab:observables}) to experimental data.
The top row has explicit model calculations at each of the 300 design points;
recall that all model parameters vary across their full ranges, leading to the large spread in computed observables.
The bottom row shows emulator predictions of 100 random samples from the identified particle posterior distribution (these are visually indistinguishable for the charged particle posterior).
Here, the model has been calibrated to experiment, so its calculations are clustered tightly around the data---although some uncertainty remains since the samples are drawn from a posterior distribution of finite width.
Overall, the calibrated model provides an excellent simultaneous fit to all observables except the pion/kaon yield ratio, which (although it is difficult to see on a log scale) deviates by roughly 10--30\%.
We address this deficiency in the following subsections.

\subsection{Initial condition parameters}

\fig{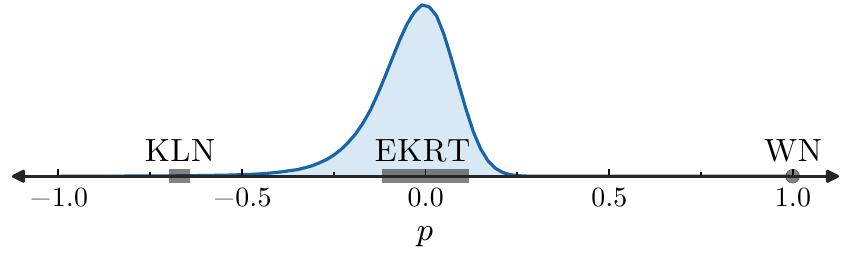}{
  Posterior distribution of the \protect\trento\ entropy deposition parameter $p$ introduced in Eq.~\eqref{eq:genmean}.
  Approximate $p$-values are annotated for the KLN ($p~\approx~0.67~\pm~0.01$), EKRT ($p~\approx~0.0~\pm~0.1$), and wounded nucleon ($p = 1$) models.
}

The first four parameters are related to the initial condition model.
Proceeding in order:

The normalization factor is not a physical parameter but nonetheless must be tuned to fit overall particle production.
Both calibrations produced narrow posterior distributions, with the identified particle result located slightly lower to compromise between pion and kaon yields.
There are some mild correlations between the normalization and other parameters that affect particle production.

The \trento\ entropy deposition parameter $p$ introduced in Eq.~\eqref{eq:genmean} has a remarkably narrow distribution, with the two calibrations in excellent agreement.
The estimated value is essentially zero with approximate 90\% uncertainty $\pm0.2$, meaning that initial state entropy deposition is roughly proportional to the geometric mean of participant nuclear thickness functions, $s \sim \sqrt{\T_A\T_B}$.
This confirms previous analysis of the \trento\ model which demonstrated that $p \approx 0$ simultaneously produces the correct ratio between initial state ellipticity and triangularity and fits multiplicity distributions for a variety of collision systems \cite{Moreland:2014oya}.
We observe little correlation between $p$ and any other parameters, suggesting that its optimal value is mostly factorized from the rest of the model.

Further, recall that the $p$ parameter smoothly interpolates among different classes of initial condition models;
Fig.~\ref{fig:posterior_p_arrows} shows an expanded view of the posterior distribution along with the approximate $p$-values for the other models in Fig.~\ref{fig:cgc_compare}.
The EKRT model (and presumably IP-Glasma as well) lie squarely in the peak---this helps explain their success---while the KLN and wounded nucleon models are considerably outside.

The distributions for the multiplicity fluctuation parameter $k$ are quite broad, indicating that it's relatively unimportant for the present model and observables.
Indeed, these fluctuations are overwhelmed by nucleon position fluctuations in large collision systems such as Pb+Pb.

The Gaussian nucleon width $w$ has fairly narrow distributions mostly within 0.4--0.6 fm.
It appears we did not extend the initial range low enough and so the posteriors are truncated;
however we still resolve peaks at ${\sim}0.43$ and ${\sim}0.49$ fm for the identified and charged particle calibrations, respectively.
Since the distributions are asymmetric, the median values are somewhat higher than the modes.
The quantitative estimates and uncertainties are in good agreement with the gluonic widths extracted from deep inelastic scattering data at HERA \cite{Chekanov:2004mw, Kowalski:2006hc, Rezaeian:2012ji} and support the values used in EKRT and IP-Glasma studies \cite{Niemi:2015qia, Schenke:2012wb}.
We also observe striking correlations between the nucleon width and QGP viscosities---this is because decreasing the width leads to smaller scale structures and steeper gradients in the initial state.
So e.g.\ as the nucleon width decreases, average transverse momentum increases, and bulk viscosity must increase to compensate.
This explains the strong anti-correlation between $w$ and $\zeta/s$ norm.

\subsection{QGP medium parameters}

The shear viscosity parameters $(\eta/s)_\text{min,slope}$ set the temperature dependence of $\eta/s$ according to the linear ansatz
\begin{equation}
  (\eta/s)(T) = (\eta/s)_\text{min} + (\eta/s)_\text{slope} (T - T_c)
  \label{eq:etas2}
\end{equation}
for $T > T_c$.
The full parametrization Eq.~\eqref{eq:etas} also includes a constant $(\eta/s)_\text{hrg}$ for $T < T_c$; this parameter was included in the calibration but yielded an essentially flat posterior distribution, implying that it has little to no effect.
This is not surprising, since hadronic viscosity is largely handled by UrQMD, not the hydrodynamic model.
Therefore, we omit $(\eta/s)_\text{hrg}$ from the posterior distribution visualizations and tables.

\begin{table}[b]
  \caption{
    \label{tab:posterior}
    Estimated parameter values (medians) and uncertainties (90\% credible intervals) from the posterior distributions calibrated to identified and charged particle yields (middle and right columns, respectively).
    The distribution for \Tsw\ based on charged particles is essentially flat, so we do not report a quantitative estimate.
  }
  \begin{ruledtabular}
    \begin{tabular}{lll}
      & \multicolumn{2}{c}{Calibrated to:} \\
      \noalign{\smallskip}\cline{2-3}\noalign{\smallskip}
      Parameter & \multicolumn{1}{c}{Identified} & \multicolumn{1}{c}{Charged} \\
      \paddedhline
      Normalization & \parbox{3.5em}{\hfill $120.$}$_{-8.}^{+8.}$ & \parbox{3.5em}{\hfill $132.$}$_{-11.}^{+11.}$ \\ \noalign{\smallskip}
$p$ & \parbox{3.5em}{\hfill $-0.02$}$_{-0.18}^{+0.16}$ & \parbox{3.5em}{\hfill $0.03$}$_{-0.17}^{+0.16}$ \\ \noalign{\smallskip}
$k$ & \parbox{3.5em}{\hfill $1.7$}$_{-0.5}^{+0.5}$ & \parbox{3.5em}{\hfill $1.6$}$_{-0.5}^{+0.6}$ \\ \noalign{\smallskip}
$w$ [fm] & \parbox{3.5em}{\hfill $0.48$}$_{-0.07}^{+0.10}$ & \parbox{3.5em}{\hfill $0.51$}$_{-0.09}^{+0.10}$ \\ \noalign{\smallskip}
$\eta/s$ min & \parbox{3.5em}{\hfill $0.07$}$_{-0.04}^{+0.05}$ & \parbox{3.5em}{\hfill $0.08$}$_{-0.05}^{+0.05}$ \\ \noalign{\smallskip}
$\eta/s$ slope [GeV$^{-1}$] & \parbox{3.5em}{\hfill $0.93$}$_{-0.92}^{+0.65}$ & \parbox{3.5em}{\hfill $0.65$}$_{-0.65}^{+0.77}$ \\ \noalign{\smallskip}
$\zeta/s$ norm & \parbox{3.5em}{\hfill $1.2$}$_{-0.3}^{+0.2}$ & \parbox{3.5em}{\hfill $1.1$}$_{-0.5}^{+0.5}$ \\ \noalign{\smallskip}
$T_\mathrm{switch}$ [GeV] & \parbox{3.5em}{\hfill $0.148$}$_{-0.002}^{+0.002}$ & \hspace{3em}--- \\ \noalign{\smallskip}

    \end{tabular}
  \end{ruledtabular}
\end{table}

Examining the marginal distributions for $\eta/s$ min and slope, we see a clear preference for $(\eta/s)_\text{min} \lesssim 0.15$ and a slight disfavor of steep slopes;
however, the marginal distributions do not paint a complete picture.
The joint distribution shows a salient correlation between the two parameters, hence, while neither $\eta/s$ min nor slope are strongly constrained independently, a linear combination is quite strongly constrained.
Figure~\ref{fig:etas_estimate} visualizes the complete estimate of the temperature dependence of $\eta/s$ via the median min and slope from the posterior (for identified particles) and a 90\% credible region.
This visualization corroborates that the posterior for $(\eta/s)(T)$ is markedly narrower than the prior and further reveals that the uncertainty is smallest at intermediate temperatures, $T \sim {}$200--225 MeV.
We hypothesize that this is the most important temperature range for the present observables at $\sqrts = 2.76$~TeV---perhaps it is where the system spends most of its time and hence where most anisotropic flow develops, for instance---and thus the data provide a ``handle'' for $\eta/s$ around 200 MeV.
Data at other beam energies and other, more sensitive observables could provide additional handles at different temperatures, enabling a more precise estimate of the temperature dependence of $\eta/s$.

This result for $(\eta/s)(T)$ supports several recent findings using other models:
a detailed study using the EKRT model \cite{Niemi:2015qia} showed that a combination of RHIC and LHC data prefer a flat or shallow high-temperature slope, while an analysis using a three-dimensional constituent quark model \cite{Denicol:2015nhu} demonstrated that a similar flat or shallow slope best describes the rapidity dependence of elliptic flow at RHIC.
In addition, the estimated temperature-averaged shear viscosity is consistent with the (constant) $\eta/s = 0.095$ reported \cite{Ryu:2015vwa} using the IP-Glasma model and the same bulk viscosity parametrization, Eq.~\eqref{eq:zetas}.
Finally, the present result remains compatible (within uncertainty) with the KSS bound $\eta/s \geq 1/4\pi$ \cite{Danielewicz:1984ww, Policastro:2001yc, Kovtun:2004de}.

\fig{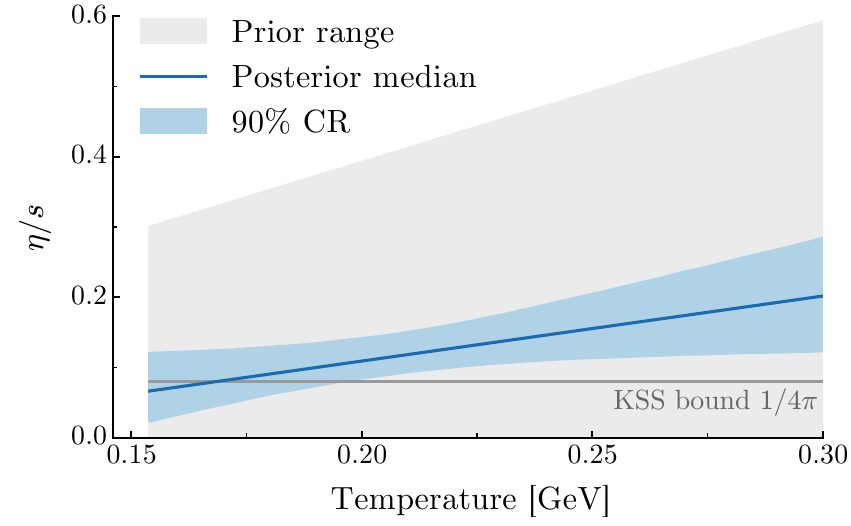}{
  Estimated temperature dependence of the shear viscosity $(\eta/s)(T)$ for $T > T_c = 0.154$ GeV.
  The gray shaded region indicates the prior range for the linear $(\eta/s)(T)$ parametrization Eq.~\eqref{eq:etas2},
  the blue line is the median from the posterior distribution,
  and the blue band is a 90\% credible region.
  The horizontal gray line indicates the KSS bound $\eta/s \geq 1/4\pi$ \cite{Danielewicz:1984ww, Policastro:2001yc, Kovtun:2004de}.
}

\fig*{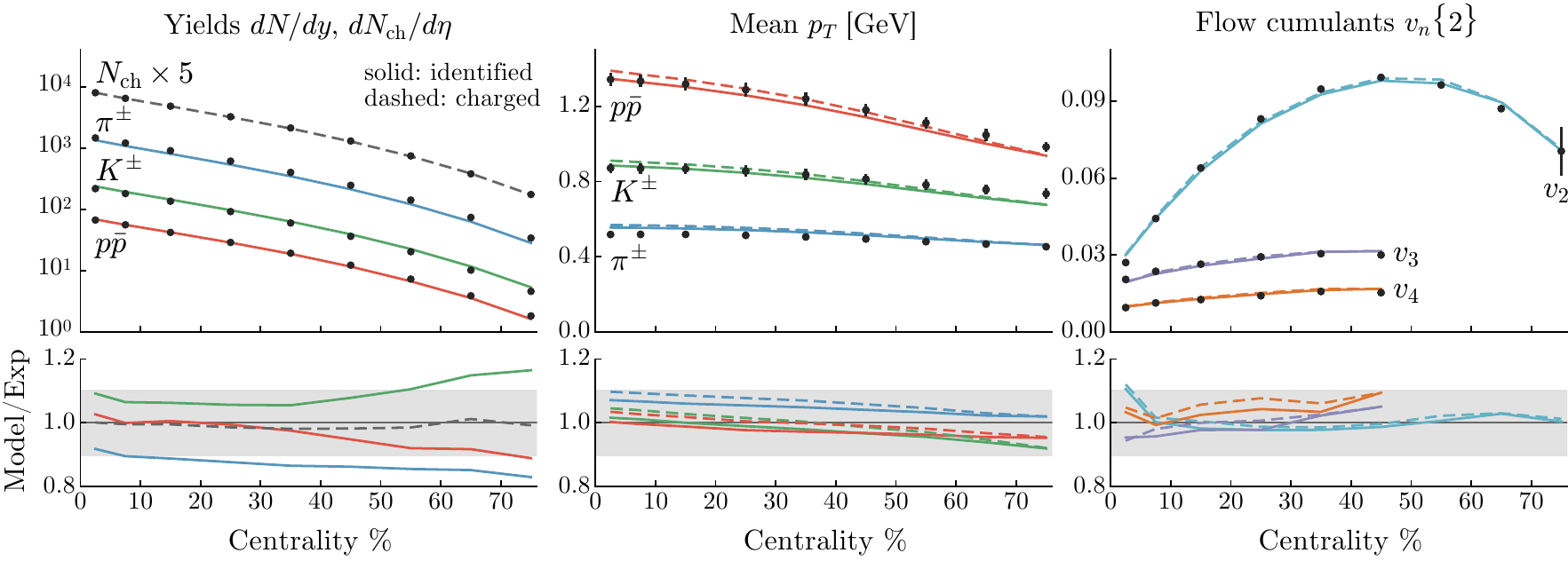}{
  Model calculations using the high-probability parameters listed in Table~\ref{tab:mode_params}.
  Solid lines are calculations using parameters based on the identified particle posterior,
  dashed lines are based on the charged particle posterior,
  and points are data from the ALICE experiment \cite{Abelev:2013vea,ALICE:2011ab}.
  Top row: calculations of identified or charged particle yields $dN/dy$ or $d\nch/d\eta$ (left), mean transverse momenta $\avg{p_T}$ (middle), and flow cumulants $\vnk n 2$ (right) compared to data.
  Bottom: ratio of model calculations to data, where the gray band indicates $\pm10$\%.
}

One should interpret the estimate of $(\eta/s)(T)$ depicted in Fig.~\ref{fig:etas_estimate} with care.
We asserted a somewhat restricted linear parametrization reaching a minimum at a fixed temperature, and evidently may not have extended the prior range for the slope high enough to bracket the posterior distribution;
these assumptions, along with the flat 10\% uncertainty [see Eq.~\eqref{eq:uncertainty}], surely affect the precise result.
And in general, a credible region is not a strict constraint---the true function may lie partially or completely (however improbably) outside the estimated region.
Yet the overarching message holds: we find the least uncertainty in $\eta/s$ at intermediate temperatures, and estimate that its temperature dependence has at most a shallow positive slope.

For the $\zeta/s$ norm [the prefactor for the parametrization Eq.~\eqref{eq:zetas}], the calibrations yielded clearly peaked posterior distributions located slightly above one.
Hence, the estimate is comfortably consistent with leaving the parametrization unscaled, as in \cite{Ryu:2015vwa}.
As noted in the previous subsection, there is a strong anti-correlation between $\zeta/s$ norm and the nucleon width.
We also observe a positive correlation with $\eta/s$ min, which initially seems counterintuitive.
This dependence arises via the nucleon width:
increasing bulk viscosity requires decreasing the nucleon width, which in turn necessitates increasing shear viscosity to damp out the excess anisotropy.
Given the previously mentioned shortcomings in the current treatment of bulk viscosity (neglecting bulk corrections at particlization, lack of a dynamical pre-equilibrium phase), we refrain from making any quantitative statements.
What is clear, however, is that a nonzero bulk viscosity is necessary to simultaneously describe transverse momentum and flow data.

The distributions for the particlization temperature $T_\text{switch}$ have by far the most dramatic difference between the two calibrations.
The posterior from identified particle yields shows a sharp peak centered at $T \approx 148$~MeV, just below $T_c = 154$~MeV;
but with charged particle yields, the distribution is nearly flat.
This is because the final particle ratios---while somewhat modified by scatterings and decays in the hadronic phase---are largely determined by the thermal ratios at the particlization temperature.
So, when we require the model to describe identified particle yields, $T_\text{switch}$ is tightly constrained;
on the other hand, lacking these data there is little else to determine an optimal switching temperature.
This reinforces the original hybrid model postulate---that both hydro and Boltzmann transport models predict the same medium evolution within a temperature window \cite{Bass:2000ib,Nonaka:2006yn,Petersen:2008dd}.

Note that, while we do see a narrow peak for $T_\text{switch}$, the model cannot simultaneously fit pion, kaon, and proton yields;
in particular, the pion/kaon ratio is 10--30\% low.
The peak thus arises from a compromise between pions and kaons---not an ideal fit---so we do not consider the quantitative value of the peak to be particularly meaningful.
This is a long-standing issue in hybrid models \cite{Song:2013qma} and therefore likely indicates a more fundamental problem with the particle production scheme rather than one with this specific model.

\subsection{Verification of high-probability parameters}

As a final verification of emulator predictions and the model's accuracy, we calculated a large number of events using high-probability parameters and compared the resulting observables to experiment.
We chose two sets of parameters based on the peaks of the posterior distributions, listed in Table~\ref{tab:mode_params}.
These values approximate the ``most probable'' parameters and the corresponding model calculations should optimally fit the data.

\begin{table}[t]
  \caption{
    \label{tab:mode_params}
    High-probability parameters chosen based on the posterior distributions and used to generate Fig.~\ref{fig:mode_observables}.
    Pairs of values separated by slashes are based on identified / charged particle yields, respectively.
    Single values are the same for both cases.
  }
  \begin{ruledtabular}
    \begin{tabular}{ll@{\hspace{2em}}ll}
      \multicolumn{2}{c}{Initial condition} & \multicolumn{2}{c}{QGP medium} \\
      \paddedhline
      norm & 120. / 129.    & $\eta/s$ min      & 0.08  \\
      $p$  & 0.0            & $\eta/s$ slope    & 0.85 / 0.75 GeV$^{-1}$  \\
      $k$  & 1.5  / 1.6     & $\zeta/s$ norm    & 1.25 / 1.10 \\
      $w$  & 0.43 / 0.49 fm & $T_\text{switch}$ & 0.148 GeV \\
    \end{tabular}
  \end{ruledtabular}
\end{table}

We evaluated \order 5 minimum-bias events (no emulator) for each set of parameters and computed observables, shown along with experimental data in Fig.~\ref{fig:mode_observables}.
Solid lines represent calculations using parameters based on the identified particle posterior while dashed lines are based on the charged particle posterior.
Note that these calculations include a peripheral centrality bin (70--80\%) that was not used in parameter estimation.

We observe an excellent overall fit; most calculations are within 10\% of experimental data, the notable exceptions being the pion/kaon ratio (discussed in the previous subsection) and central elliptic flow, both of which are general problems within this class of models.
Total charged particle production is nearly perfect---within 2\% of experiment out to 80\% centrality---indicating that the issues with identified particle ratios arise in the particlization and/or hadronic phases, not in initial entropy production.
The $v_2$ mismatch in the most central bin is a manifestation of the experimental observation that elliptic and triangular flow converge to nearly the same value in ultra-central collisions \cite{ALICE:2011ab, CMS:2013bza}, a phenomenon that hydrodynamic models have yet to explain \cite{Denicol:2014ywa,Shen:2015qta}.

\section{Summary and conclusions}

We have used Bayesian methodology to quantitatively estimate initial condition and transport properties of the QGP medium produced in relativistic heavy-ion collisions.
We coupled a parametric initial condition model to viscous hydrodynamics and a hadronic afterburner, calibrated the full model to a variety of bulk observables, and established a number of salient constraints on model parameters, including a relation between the minimum value and slope of the temperature-dependent shear viscosity, a clear signal for a nonzero bulk viscosity, and a robust constraint on initial state entropy deposition.

The parametric initial condition model used in this analysis, \trento, smoothly interpolates among various physically reasonable entropy deposition schemes, ranging from a wounded nucleon model to specific calculations in color glass condensate effective field theory.
This flexibility is ideal for model-to-data comparison, since it allows the analysis framework to optimize the initial conditions with minimal theoretical assumptions.

The heavy-ion collision transport dynamics were simulated using an event-by-event hybrid model with viscous hydrodynamics for the early hot and dense stage and a microscopic hadronic afterburner for the later dilute stage.
The hydrodynamic model uses a modern continuum extrapolated lattice equation of state and implements temperature-dependent shear and bulk viscous corrections.
To constrain the viscosities, we parametrized their temperature dependence with several tunable model parameters for optimization.

With the full evolution model in hand, we applied Bayesian methods to estimate its various input parameters.
We evaluated the model at several hundred points in parameter space, calculated bulk observables at each point, and trained a Gaussian process emulator to interpolate the model calculations.
Then, we used a Markov chain Monte Carlo (MCMC) algorithm to systematically explore parameter space---with the emulator acting as a stand-in for the complete model---and calibrate the model to optimally reproduce experimental data, thereby extracting posterior probability distributions for all parameters and their correlations.

The primary results of this work are the posterior distributions, shown in Fig.~\ref{fig:posterior}, and the
corresponding quantitative estimates of each parameter, presented in Table~\ref{tab:posterior}.
These distributions contain a wealth of information about QGP initial condition and medium properties; here we summarize the key features:
\begin{enumerate}[itemsep=0pt, leftmargin=2\parindent]
  \item
    Based on the \trento\ initial condition parametrization, we find that initial entropy deposition is approximately proportional to the geometric mean of local participant nuclear densities.
    This scaling is functionally similar to the notably successful EKRT and IP-Glasma models.
  \item
    The preferred Gaussian nucleon width is roughly $0.5 \pm 0.1$~fm, consistent with values extracted from HERA deep inelastic scattering data.
  \item
    For the temperature-dependent specific shear viscosity $(\eta/s)(T)$, we asserted a linear parametrization reaching its minimum at the QCD phase transition temperature.
    The data cannot individually constrain both the minimum value and the slope, but do constrain a linear combination, as shown in Fig.~\ref{fig:etas_estimate}.
    The uncertainty on $\eta/s$ is smallest at intermediate temperatures, $T \sim {}$200--225 MeV;
    we hypothesize that this is the most important temperature range at $\sqrts = 2.76$~TeV,
    and that including data from additional beam energies would enable a more precise estimate of $(\eta/s)(T)$.
  \item
    We observe a clear preference for a nonzero bulk viscosity, which is necessary to simultaneously describe transverse momentum and flow data.
    We refrain from making any quantitative statements given current limitations in the treatment of bulk viscosity.
  \item
    The result for the particlization temperature (when the model switches from hydrodynamics to hadronic afterburner) depends strongly on the observables used for calibration.
    When fitting to identified pion, kaon, and proton yields, the temperature is tightly constrained just below the QCD transition temperature.
    On the other hand, when the identified yields are replaced with total charged particle yields, there is essentially no preference within the considered range.
    This implies that both stages of the hybrid model simulate the same medium evolution near the QGP transition, but not the same hadronic chemistry.
\end{enumerate}

The aforementioned parameter estimates allow us to assess the performance of a systematically optimized model.
To this end, we evaluated the full model using high-probability parameters based on the posterior distributions.
The resulting charged particle yields, mean transverse momenta, and flow cumulants agree with experiment at the 10\% level, as shown in Fig.~\ref{fig:mode_observables}.

In future work, we plan to include data from multiple beam energies---we anticipate that a combined analysis of data at $\sqrts = 200$ GeV, 2.76 TeV, and 5.02 TeV will enable a precise extraction of temperature-dependent QGP transport coefficients.
We will also consider new, sensitive observables such as correlations between flow harmonics of different order.

We will implement several improvements to the physical models, including a free streaming stage for pre-equilibrium dynamics and bulk viscous corrections at particlization.
These changes will especially improve estimates of the specific bulk viscosity $\zeta/s$.

Finally, we plan to improve the treatment of experimental and model uncertainties, essential for rigorous quantitative uncertainties on estimated parameters.

\vspace*{\baselineskip}

\newcommand{\nicelink}[2][http]{\mbox{\href{#1://#2}{\nolinkurl{#2}}}}

All code used in this study is publicly available:
the \trento\ initial condition model at \nicelink{qcd.phy.duke.edu/trento},
the iEBE-VISHNU package at \url{u.osu.edu/vishnu},
UrQMD at \url{urqmd.org},
the workflow for generating events at \nicelink[https]{github.com/jbernhard/heavy-ion-collisions-osg},
and the source for this manuscript including all figures and tables at \nicelink[https]{github.com/Duke-QCD/trento-paper-2}.

\begin{acknowledgments}
The authors thank Scott Pratt, Berndt M\"uller, Harri Niemi, Bj\"orn Schenke, and Ron Soltz for helpful discussions and clarifications.
JL and UH gratefully acknowledge many clarifying discussions with Chun Shen during the implementation of bulk viscosity into VISH2+1.
This research was completed using 1.5 million CPU hours provided by the Open Science Grid \cite{Pordes:2007zzb,Sfiligoi:2010zz}, which is supported by the National Science Foundation and the U.S.\ Department of Energy's Office of Science.
SAB and JEB are supported by the U.S.\ Department of Energy Grant no.~DE-FG02-05ER41367,
JSM by the DOE/NNSA Stockpile Stewardship Graduate Fellowship under Grant no.~DE-FC52-08NA28752,
and JL and UH by the U.S.\ Department of Energy, Office of Science, Office for Nuclear Physics Office under Award DE-SC0004286.
\end{acknowledgments}

\bibliography{trento2}

\end{document}